
%
\magnification=\magstep1
\vsize=23truecm
\hsize=15.5truecm
\hoffset=.2truecm
\voffset=.8truecm
\parskip=.2truecm

\font\ti=cmbx10 scaled\magstep1
\font\eightrm=cmr8
\font\ninerm=cmr9
\def\br{\hfill\break\noindent}

\def \es{e^{-\kappa \sigma}}
\def \ot {\otimes}
\def \g5{\gamma_5}
\def \ra{\rightarrow}
\def \l4{\Bigl( {\rm Tr}( KK^*)^2-({\rm Tr}KK^*)^2\Bigr)}

\def \k2{{\rm Tr}KK^*}
\def \slash#1{/\kern -6pt#1}

\pageno=0

%
%
\baselineskip=.5truecm
\footline={\hfill}
{\hfill ZU-TH- 16/1993}
\vskip.1truecm
{\hfill June 1993}
\vskip2.1truecm
\centerline{\ti Constraints on the Higgs and Top Quark Masses}
\centerline{\ti From Effective Potential and Non-Commutative Geometry }
\vskip1.2truecm
\centerline{  A. H. Chamseddine$^{1,2}$ \footnote*
{\ninerm Supported in part by the Swiss National Foundation (SNF)}
and J. Fr\"ohlich$^{3}$ \footnote{\dag}{\ninerm Permanent address:
Theoretische Physik, ETH, CH 8093 Z\"urich Switzerland}}
\vskip.8truecm
\centerline{$^{1}$ Theoretische Physik, Universit\"at Z\"urich, CH
8001 Z\"urich Switzerland}
\centerline{$^{2}$ Universit\'e Aix-Marseille II, Luminy, France }
\centerline{$^{3}$ Institut des Hautes Etudes Scientifiques,
F-91440 Bures-sur-Yvette, France}
\vskip1.2truecm

\centerline{\bf Abstract}
\vskip.5truecm

\noindent
We consider the standard model in the formulationo of
non-commutative geometry, for a Euclidean space-time consisting of
two copies. The electroweak scale is set by the vacuum
expectation value of a scalar field and is undetermined at
the classical level. By adding the Coleman-Weinberg effective
potential, that scale turns out to be fixed. Provided that the
renormalized form of the Lagrangian maintains the vanishing
of the cosmological constant, we show that the only solutions
for the minimization equations
of the total potential occur in the narrow band $146.2 \le
 m_t \le 147.4 $ Gev for the top quark mass, with
the corresponding Higgs mass $117.3 \le  m_H\le 142.6$
Gev.
\vfill

\eject
\baselineskip=.6truecm
\footline={\hss\eightrm\folio\hss}
Connes' [1] framework of non-commutative geometry provides
 a geometrical
interpretation of the Higgs field necessary to break the gauge
symmetry spontaneously. The structure of space-time is taken to
be a product of a continuous four-dimensional Riemannian manifold
times a discrete set of two points. For such a structure of
space-time
the usual methods of differential geometry fail and must be
replaced with the more general framework of non-commutative geometry.
Using the non-commutative setting, Connes and Lott [2] recovered the
standard model with all its parameters, at the classical level.
At present, it is not known how to quantize the non-commutative
action directly. We are left only
with the possibility of quantizing the resulting theory in the
usual way. The quantum corrections are then  given by familiar
expressions. There is, however, one important difference
between this approach and the standard analysis connected with
the gravitational effects of the discrete geometry.
This have, under certain conditions, very surprising consequences.
Explaining and exploiting these effects is the main concern
of this note.
Geometrically, the distance between the two copies of the
four-dimensional Minkowski space is the inverse of the electoweak
scale. At smaller scales the flat manifold is curved and the
distance between the two copies becomes a dynamical scalar field
[3]. The leptonic Dirac
operator associated with Connes-Lott space takes the form
$$
D_l=\pmatrix{\gamma^a e_a^{\mu} (\partial_{\mu} +\ldots )\ot 1_2
\ot 1_3&\g5 \es \ot M_{12}\ot k\cr
\g5 \es \ot M_{12}^* \ot k^*&\gamma^a e_a^{\mu}(\partial_{\mu} +\ldots
)\ot 1_3\cr},\eqno(1)
$$
where $M_{12}=\mu \pmatrix{0 \cr 1 \cr}$, $k$ is a $3\times 3$
family mixing matrix, $\kappa^{-1}$ is
the Planck scale, and the dots correspond to the spin-connection
and derivatives of the $\sigma $ field. It was shown in [3] that the
gravitational action associated with this geometry yields the
Einstein-Hilbert action of the metric $g_{\mu \nu}=e_{\mu}^a e_{\nu}^b
\eta_{ab}$, as well as a kinetic action for the scalar field $\sigma $.
Comparing the Dirac operator (1) with the one used in constructing
the standard model [2] we find that we must identify $H_0=<\mu \es >$,
where $H_0$ is the vacuum expectation value (vev) of the Higgs field.
In the gravitational part of the action, the field $\sigma $
appears only through
derivatives and, in the Yang-Mills-Higgs action, it appears as a
scale factor, in powers of $\es $. There is no pure potential
term for $\sigma $, and its vev is
undetermined at the classical level.  We shall show
how $\es $ may acquire a vev as a consequence of radiative
corrections,  and how this imposes severe constraints on the top
quark mass and the Higgs mass.

The non-commutative construction usually produces models with a
restricted set of parameters [2-4]. This is, however, not the case
for the standard model where
the same number of parameters is obtained as in the usual approach.
To determine explicitly
the parameters of the theory and the $\sigma $ couplings, we indicate
briefly, and in a simplified way, the steps of [2]
leading to the classical action.
The leptons are arranged in three copies of the multiplet
$L=\pmatrix{\nu_L \cr e_L^-\cr e_R^-\cr}$ with a Dirac
operator $D_l $ of a form given by equation (1).
To incorporate the quarks correctly, a bimodule with algebras
$\cal A$ and $\cal B$ is introduced [2]. When acting on the leptonic
Hilbert space, the  elements $a^i,b^i,\cdots $ of the algebra $\cal A $
describing the non-commutative space have a representation
:$a\ra {\rm diag }(a_1,a_2)$,
where $a_1$ is a $2\times 2$ quaternionic matrix of continuous functions
and $a_2$ is a continuous function. A one-form  in
$\Omega^1(\cal A)$ is given by $\rho =\sum_i a^idb^i$,
and the involutive map $\pi_l $ is defined by
$$\pi_l (\rho )=\sum_i a^i[D_l,b^i],
$$
and this is easily evaluated to be
$$
\pi_l (\rho )=\pmatrix{ A_1 & \g5 k \es H\cr \g5 k \es H^* & A_2
\cr}, \eqno(2)
$$
where  $A_1$ and $A_2$ are
the U(2) and U(1) gauge fields. The algebra $\cal B$ is taken to
be $M_1(C)\oplus M_3(C)$ commuting with the action of $\cal A$,
and the mass  matrices in the Dirac
operator taken to be zero when acting on the elements of $\cal B$.
Then the one-form $\eta $ in $\Omega^1(\cal B)$ has the simple form
$\pi_l (\eta )=B_1 {\rm diag}(1_2,1)$. The leptonic
action is
$$
<L,(D+\rho +\eta )L>=\int d^4 x\overline L (D_l +\pi_l (\rho )
+\pi_l(\eta ))L.
\eqno(3)
$$
The leptonic part of the electroweak bosonic action is
$$
I_l = Tr_w \Bigl( C_l (\theta_{\rho}
+\theta_{\eta})^2 D_l^{-4}+ \Bigr),  \eqno(4)
$$
where $Tr_w $ is the Dixmier trace [2], $C_l$ is a constant element
of $\cal A $, and $\theta_{\rho} =d\rho +\rho^2 $ is the curvature.
Similarly, the quarks are arranged in three copies of the
multiplet $Q=\pmatrix{u_L \cr d_L \cr d_R \cr u_R}$. The elements
of the algebra $\cal A $ have the representation $a\ra {\rm diag}
(a_1, a_2,\overline {a_2})$ where $a_1$ is a $2\times 2$
quaternionic matrix
of continuous functions and $a_2$ is a complex valued continuous function.
The Dirac operator $D_q$ associated
with this representation is
$$
D_q=\pmatrix{\gamma^a e_a^{\mu} (\partial_{\mu} +\ldots )\ot 1_2
\ot 1_3 &\g5 \es \ot M_{12} \ot k{'}&\g5 \es \ot \tilde{M_{12}}\ot k{''}\cr
\g5 \es \ot M_{12}^* \ot k^{'*}&\gamma^a
e_a^{\mu}(\partial_{\mu} +\ldots )
\ot 1_3 &0\cr
\g5 \es \ot \tilde {M_{12}}^*\ot k^{''*}&0&\gamma^a e_a^{\mu}
(\partial_{\mu} +\ldots )\ot 1_3}, \eqno(5)
$$
where $k'$ and $k^{''}$ are $3\times 3$ family mixing matrices,
and $\tilde {M_{12}}=\mu \pmatrix{1\cr 0\cr}$.
Then the one-form in $\Omega^1(\cal A)$ has the representation
$$
\pi_q(\rho )=\pmatrix{A_1\ot 1_3& \g5 H\ot k'&\g5 \tilde H\ot
k^{''}\cr \g5 H^*\ot k^{'*}&A_2\ot 1_3&0\cr
\g5 {\tilde H}^*\ot k^{''*}&0&\overline {A_2}\cr},\eqno(6)
$$
where $\tilde H_a=\epsilon_{ab}H^b$.
On the algebra $\cal B$ the Dirac operator has zero mass matrices,
and the one form $\eta $ in $\Omega^1(\cal B) $ has the
representation $\pi_q(\eta )= B_2{\rm diag}(1_2,1,1)$.
Imposing the unimodularity condition on the algebras $\cal A$
and $\cal B$ relates the U(1) factors in both algebras:
${\rm tr}(A_1)=0$, $A_2=B_1=-{\rm tr}B_2={i\over 2}g_1B$.
We can then write
$$\eqalign{
A_1&=-{i\over 2}g_2 A^a\sigma_a \cr
B_2&=-{i\over 6}g_1B -{i\over 2}g_3 V^i\lambda_i \cr}, \eqno(7)
$$
where $g_1,g_2,g_3$ are the U(1), SU(2), and SU(3) coupling
constants, and $\sigma^a $ and $\lambda^i $ are the Pauli
and Gell-Mann matrices respectively.
The quark part of
the electroweak bosonic action is
$$
I_q= Tr\Bigl( C_q (\theta_{\rho}^2 +\theta_{\eta}^2)D_q^{-4} \Bigr). \eqno(8)
$$
By writing $C_l ={\rm diag}(c_1,c_1,c_2)$ and $C_q={\rm diag}
(c_3,c_3,c_4,c_4)$, the bosonic action depends on the
constants $c_1,c_2,c_3,c_4,g_1,g_2,g_3$ as well as on the Yukawa
couplings and on $\es $. Normalizing the kinetic energies of the
$SU(3)$, $SU(2)$ and $U(1)$ gauge fields fixes three of the constants
$c_1,\ldots ,c_4$ in terms of $g_1,g_2,g_3$. In the special case
when $c_1=c_2=c_3=c_4$, one gets a constraint on the gauge
coupling constants as well as fixed values for the Higgs mass
and top quark mass. These relations cannot be maintained
after quantization, as can be seen from the renormalization
group equations for the coupling constants and the masses [5].
We shall not assume any such relations among the $c's$.
The Higgs sector
is then parametrized in terms of two parameters $\lambda $ and
$m$ which are functions of
of the $c's$, $k$, $k'$, $k''$ and $<H_0>$.
The bosonic part of the standard model is
$$\eqalign{
L_b &= -{1\over 4} \Bigl ( F_{\mu\nu}^3 F^{\mu\nu 3}
+F_{\mu\nu}^2 F^{\mu\nu 2} +F_{\mu\nu}^1 F^{\mu\nu 1} \Bigl) \cr
&\qquad +D_{\mu} (H+M_{12})^* D_{\nu}(H+M_{12}) g^{\mu\nu}e^{-2\kappa
\sigma} \cr
&\qquad -{\lambda \over 24} \Bigl\vert \vert H+M_{12}\vert^2
-\vert M_{12}\vert^2 \Bigr\vert^2 e^{-4\kappa \sigma} \cr}.\eqno(9)
$$
The cosmological constant comes out to be zero, naturally, at the
classical level.
The $\sigma $ dependence in (9) is a consequence of the "Weyl
invariance" of the actions (4) and (8)
under the rescaling of the Dirac operator
$D\ra e^{-w}D$, as this implies $g_{\mu\nu}\ra e^{2w}g_{\mu\nu}$ and
$\kappa\sigma \ra \kappa \sigma +w$. This can be easily seen from
the scalings: $\pi (\rho )\ra e^{-w}\pi (\rho )$ and
$\pi (\theta )\ra e^{-2w}\pi (\theta )$.
By redefining
$H+M_{12}\ra e^{\kappa\sigma} H$, the $H$ dependent terms in
(9) become
$$\eqalign{
D_{\mu}H^* D^{\mu}H +\kappa \partial_{\mu}(H^*H)\partial^{\mu}\sigma
+\kappa^2 H^*H \partial_{\mu}\partial^{\mu}\sigma
 -{\lambda \over 24}\Bigl\vert (H^*H)^2-\mu^2 e^{-2\kappa
\sigma}\Big\vert^2 \cr}.\eqno(10)
$$
The potential in (10) could be rewritten in the familiar form
$$
V_0={\lambda \over 24} (H^*H)^2 -{1\over 2}m^2 (H^*H) +{3\over 2\lambda}
m^4 ,\eqno(11)
$$
where we have set $m^2 ={\lambda \mu^2\over 6}e^{-2\kappa\sigma}$,
so that $m$ is now a field and not just a parameter. A similar
potential has also been considered in [6] with a different motivation.
The potential $V_0$ is of the same form as that of the standard
model.
At the electroweak scale, which is much smaller
than the gravitational Planck scale, the Lagrangian we consider
is renormalizable. The cosmological constant in the standard
model can be tuned to zero but in the non-commutative construction
it automatically comes out to be zero at the tree level.
We therefore assume that,
after renormalization, the bosonic action takes the same form
as $I_l+I_q$. We  warn the reader that, although this
assumption appears to be  reasonable,
we cannot prove that the most general form of a non-commutative
Yang-Mills action is preserved at the quantum level,
in the absence of some understanding of
its symmetries.  We shall proceed in our analysis on the
basis of this assumption.

Let $\phi $ be the component of the Higgs field that develop
a vev. We are then mainly interested in the potential
$$
V_0={\lambda \over 24} \phi^4 -{1\over 2}m^2\phi^2 +{3\over 2\lambda}
m^4 . \eqno(12)
$$
Minimizing with respect to $\phi $ and $m$ yields the equations
$$\eqalignno{
0&={\lambda \over 6}\phi^3 -m^2\phi , &(13)\cr
0&=-m\phi^2 +{6\over \lambda}m^3 . &(14)\cr}
$$
Both equations, (13) and (14), have the same asymmetric phase
$$
\phi^2 ={6\over \lambda}m^2 , \eqno(15)
$$
and the weak scale, $\es $, is undetermined at the classical level.
The quantum corrections to the potential are given, in the one-loop
approximation, by the
effective Coleman-Weinberg [7] potential of the standard model [8]:
$$\eqalign{
V_1&= {1\over 16\pi^2}\Bigl( {1\over 4}H^2 (\ln {H\over
M^2}-{3\over 2}) +{3\over 4}G^2 (\ln {G\over M^2} -{3\over 2})
+{3\over 2 }W^2 (\ln {W\over M^2}-{5\over 6})\cr
&\qquad \qquad +{3\over 4}Z^2 (\ln {Z\over M^2}-{5\over 6})
-3T^2 (\ln {T\over M^2}-{3\over 2})\Bigr) \cr},\eqno(16)
$$
where
$$\eqalign{
H&=-m^2 +{1\over 2} \lambda \phi^2 ,
\qquad G=-m^2 +{1\over 6} \lambda \phi^2 ,\cr
W&={1\over 4}g_2^2 \phi^2 ,\qquad Z={1\over 4}(g_2^2 +g_1^2)\phi^2 ,
\qquad T={1\over 2}h^2 \phi^2, \cr}\eqno(17)
$$
and $M$ is the renormalization scale. At the classical minimmum
$\phi^2 ={6\over \lambda }m^2 $, $H,W,Z$ and $T$ are respectively,
the squares
of the masses of the Higgs, $W^{\pm}$, Z and t particles.
The potential  $V_1$ is independent of $M$ because the coupling
constans $g_1,g_2,h,$ and $\lambda $ depend on $M$ through the
renormalization group equations in such a way that
${\partial V_1 \over \partial M}=0$.

Minimizing the total potential $V_0+V_1$ with respect to the
fields $\phi $ and $m$ gives respectively
$$\eqalignno{
0&=\phi \Bigl( G+{1\over 32\pi^2}\bigl( \lambda H(\ln {H\over M^2}-1)
+\lambda G(\ln {G\over M^2}-1)-12h^2T(\ln {T\over M^2}-1) \cr
&\qquad \qquad +3g_2^2 W(\ln {W\over M^2}-{1\over 3})
+ {3\over 2}(g_2^2+g_1^2)Z(\ln {Z\over M^2}-{1\over 3})
\bigr)\Bigr), &(18)\cr
0&=-m\Bigl( {6\over \lambda }G +{1\over 16\pi^2} \bigl(
H(\ln {H\over M^2}-1) +3G(\ln {G\over M^2}-1)\bigr).\Bigr)&(19)\cr}
$$
At the scale $M=m_Z$, the mass of the Z-particle, the coupling
constants $g_1,g_2$ as well as the vev $\phi $ are known
from experimental data, corrected with the help of the renormalization
group equations [8]:
$$
g_2=0.650, \qquad g_1=0.358 ,\qquad \phi =246 \ {\it Gev}, \eqno(20)
$$
and this implies that
$$
W=m_W^2=6392.002 \ {\it Gev}^2, \qquad Z=m_Z^2=8330.996
\ {\it Gev}^2 .\eqno(21)
$$
The only unknowns in the minimization equations are $\lambda $,
$m$ and the square of the top quark mass $T=m_t^2$. (In reality,
$T$ is the sum of the squares of all the quark masses, but this
is dominated by the top quark mass). We shall use the top quark
mass as a parameter and solve equations (18) and (19) for the full
range of this parameter. We first express $\lambda $ and $m^2$
in terms of $H$ and $G$:
$$
\lambda={3\over \phi^2}(H-G),\qquad m^2={1\over 2}(H-3G), \eqno(22)
$$
and $h$ in terms of $T$: $h^2={2T\over \phi^2 }$. After rescaling
$$
G=\overline G M^2,\qquad H=\overline H M^2, \qquad T=\overline T M^2,
\eqno(23)
$$
the asymmetric solution of equations (18) and (19) is given by the
solution of the following two equations:
$$
\eqalignno{
0 &=\overline G + {M^2\over 32\pi^2 \phi^2}(\overline H
-\overline G)\Bigl( \overline H (\ln \overline H -1)+3\overline G
(\ln \overline G -1)\Bigr), &(24) \cr
0 &=\overline G +{3M^2\over 32\pi^2\phi^2}(\overline H-\overline G)
\Bigl( \overline H (\ln \overline H -1)+\overline G (\ln \overline
G -1)\Bigr) -{g_2^2+g_1^2\over 64\pi^2}\cr
&\qquad +{3g_2^4\phi^2\over 128\pi^2M^2}\Bigl( \ln {g_2^2\phi^2
\over 4M^2}-{1\over 3}\Bigr) -{3M^2\over 4\pi^2\phi^2}
\overline T^2 (\ln \overline T-1). &(25) \cr}
$$
These equations, being complicated functions of $\overline H$ and
$\overline G$, could only be solved numerically, for various values
of $\overline T $. The numerical solutions were easily obtained using
{\it Mathematica}. Before presenting the
solutions, we note that for a given value of $\overline H,
\overline G,$ and $\overline T$, the Higgs mass can be determined
from the formula $m_H^2={\partial^2 V\over \partial \phi^2}$ which
gives
$$\eqalign{
m_H^2&=M^2\Bigl( (\overline H-\overline G)+{9M^2\over 16\pi^2
\phi^2}(\overline H-\overline G)^2(\ln \overline H +{1\over 3}
\ln \overline G )\cr
&\qquad \qquad +{3g_2^4\phi^2\over 64\pi^2 M^2}\ln {g_2^2\phi^2\over 4M^2}
\ -{3M^2\over 2\pi^2\phi^2}\overline T^2 \ln \overline T
\Bigr). \cr}\eqno(26)
$$
We now quote the results: There are only two classes of solutions,
for $\overline G\ll \overline H$ and for $\overline H\ll \overline G$.
In the first case we find that there are only two narrow bands
for the top quark mass where  solutions exist. The first band is
$$
0.365\le \overline T \le 0.455, \qquad \overline G\ll \overline H,
\eqno(27)
$$
corresponding to a top quark mass
$ 54.90\le m_t\le 61.35 $ Gev  which is already ruled out
experimentally. The second band is very narrow:
$$
2.57\le \overline T\le 2.61, \qquad \overline G\ll \overline H,
\eqno(28)
$$
corresponding to the top quark mass
$$
146.23\le m_t\le 147.37 \ {\it Gev},  \eqno(29)
$$
and a Higgs mass $117.26\le m_H\le 142.61$ Gev. Clearly this band
of values for the top quark mass lies within the present experimental
average of [9]
$$
m_t=149 +\pmatrix{+21\cr -47\cr} \ {\it Gev}.\eqno(30)
$$
These solutions are presented in Tables 1 and 2.
The second class of solutions occurs when
$$
1.30\le \overline T\le 2.61, \qquad \overline H\ll \overline G,
\eqno(31)
$$
corresponding to the top quark mass $104.07\le m_t\le 147.48 $ Gev,
and a Higgs mass $1208\ge m_H\ge 1197 $ Gev. These solutions
are given in Table 3. However, since
$\overline H\ll\overline G$, and since the coupling constant $\lambda
=O(-100)$, the potential, in this domain, becomes unbounded from below,
signaling the break down of the perturbative region. Requiring
stability of the electroweak potential excludes this solution.
Therefore the only accptable solution is (25) which is remarkably
constrained, considering the wide range of possibilities that one
might have apriori.

To have a better understanding of the solutions obtained, we
rewrite eqs (24), (25) using the numerical values (20), (21):
$$
(\overline H-\overline G)\overline H (\ln \overline H-1)
=12\overline T^2(\ln \overline T-1)+3.113136, \eqno(32)
$$
$$
\overline G+4.3589\time 10^{-4}(\overline H -\overline G)
\bigl( \overline H (\ln \overline H-1)+3\overline G(\ln
\overline G-1)\bigr)=0 .\eqno(33)
$$
The right-hand side of eq (32) is zero at $\overline T
=0.357644$ and at $\overline T=2.61727$, becomes negative for
the values of $\overline T$ in between and positive otherwise.
{}From eq (33) we see that
if the right-hand side of eq (32) is negative, then
$\overline G \equiv O(10^{-4})$ and  $\overline G\ll \overline
H$. Ignoring $\overline G$ in eq (32), the left-hand side of this
equation becomes negative for $0<\overline H<e$, with a minimum
value of $-{e\over 2} $ at $\overline H=\sqrt e $. The
right-hand side of eq (32) is larger than $-{e\over 2}$, for
$2.569\le \overline T\le 2.617$ and $0.357\le\overline T\le
0.457$. The numerical solutions, taking $\overline G$
into account, are shown in Tables 1 and 2. When the right-hand
side of eq (32) is positive then $\overline H >\overline G$ and
$\overline H>e$, or $\overline H<\overline G$ and $\overline H
<e$. In the first case, since $\overline G$ must be positive
for the potential not to become complex, one finds that there
are no solutions. In the second case  solutions would exist
if $\overline H\ll \overline G$, giving rise to large negative
coupling constant $\lambda $, and therefore, physically unacceptable.
These solutions are shown in Table 3.

We note that the field $\sigma $ becomes massive with the
square of the mass given by: $m_{\sigma}^2= {\partial^2 V\over \partial
\sigma^2} $ and this is equal to
$$
m_{\sigma }^2 =\kappa^2 m^2 \Bigl( 2\phi^2 {\overline
H-4\overline G\over \overline H-\overline G} +{M^2\over 16\pi^2}
\bigl( \overline H (1-\ln \overline H)+3\overline G (1-\ln
\overline G)\bigr)\Bigr). \eqno(34)
$$
For the physically acceptable solutions we have $\overline H=O(1)$,
$\overline G=O(10^{-4})$ and $m^2=O(M^2)$. Then we find from eq (29)
that
$$
m_{\sigma }^2=O(\kappa^2 M^4) , \eqno(35)
$$
so that $m_{\sigma} =O(10^{-15})$ Gev, which is unobservable.

To summarize, we have shown that the only acceptable solutions
for the minimization of the total potential exist in the
narrow band (25) of the top quark mass and the Higgs mass.
Of course, these predictions have, at best, heuristic value,
since the problem of fixing the form of the cosmological
constant at the one-loop level by imposing natural geometrical
constraints is not understood. However, they do suggest
that gravitational effects may play a role in understanding
masses of fermions and Higgses and that methods of non-commutative
geometry may be useful in making progress on these problems.

\vskip1truecm
{\bf\noindent Acknowledgments}\hfill\break
We would like to thank D. Wyler for very useful discussions,
and Z. Trocsanyi for help in computer programming. A.H.C would
like to thank D. Kastler for illuminating discussions,
the Universit\'e Aix Marseille II and the Centre
de Physique Th\'eorique Marseille, where part of this work
was done, for hospitality.
\vfill
\eject

\vfill
\eject

{\bf \noindent References}
\vskip.2truecm
\item{[1]} A. Connes, {\sl Publ. Math. IHES} {\bf 62} 44 (1983);\br
A. Connes, in {\sl the interface of mathematics
and particle physics }, Clarendon press, Oxford 1990, Eds
D. Quillen, G. Segal and  S. Tsou.

\item{[2]} A. Connes and J. Lott,{\sl Nucl.Phys.B Proc.Supp.}
{\bf 18B} 29 (1990), North-Holland, Amsterdam;
{\sl  Proceedings of
1991 Summer Cargese conference} p.53  editors J. Fr\"ohlich
et al (1992),Plenum Pub.;\br
For a detailed account see: D. Kastler , Marseille preprints
CPT-91/P.2610, CPT-91/P.2611 and forthcoming book.

\item{[3]} A.H. Chamseddine, G. Felder and J. Fr\"ohlich,
{\sl Zurich preprint}  ZU-TH-30\/ 1992, to appear in
{\sl Comm.Math.Phys}.

\item{[4]} R. Coquereaux, G. Esposito-Far\'ese, G. Vaillant,
{\sl Nucl. Phys.}{\bf B353} 689 (1991);\br
M. Dubois-Violette, R. Kerner, J. Madore, {\sl J. Math.
Phys.}{\bf 31} (1990) 316;\br
R. Coquereaux, G. Esposito-Far\'ese and F. Scheck,
{\sl Int.J.Mod.Phys.} {\bf A7} (1992) 6555;\br
B. Balakrishna, F. G\"ursey and K. C. Wali, {\sl Phys. Lett.}
{\bf 254B} (1991) 430;\br
A.H. Chamseddine, G. Felder and J. Fr\"ohlich, {\sl Nucl.Phys.}
{\bf B395} (1993) 672.

\item{[5]} E. Alvarez, J.M. Garcia-Bondia and C.P. Martin,
{\sl Madrid preprint}, May 1993.

\item{[6]} W. Buchm\"uller and D. Wyler, {\sl Phys.Lett}
{\bf B249} (1990) 281.

\item{[7]} E. Weinberg and S. Coleman, {\sl Phys.Rev}{\bf D7}
(1973)1888.

\item{[8]}M. Sher, {\sl Phys.Rep}{\bf 179}(1989) 274 and
references therein;\br
C. Ford, D.R.T. Jones, Y.W. Stephenson and M.B. Einhorn,
{\sl Liverpool preprint} {\bf LTH 288}.

\item{[9]} U. Amaldi, to appear in {\sl Proceedings of Salamfest
},Trieste, Italy, (March 1993), editor S. Randjbar-Daemi,
World-Scientific, Singapore.

\vfill
\eject
\vbox{\tabskip=0pt \offinterlineskip
\def\tablerule{\noalign{\hrule}}
\halign to400pt{\strut#& \vrule#\tabskip=1em plus2em&
   \hfil#& \vrule#& \hfil#& \vrule#& \hfil#&
   \vrule#& \hfill#& \vrule#& \hfil#& \vrule#
   \tabskip=0pt\cr\tablerule
&&\multispan9\hfil Table 1 \quad ($\overline G \ll \overline H $)
 \hfil&\cr\tablerule
&&\omit\hidewidth $\overline T$\hidewidth&&
  \omit\hidewidth $\overline G$\hidewidth&&
  \omit\hidewidth $\overline H$\hidewidth&&
  \omit\hidewidth $m_t$(Gev) \hidewidth&&
  \omit\hidewidth $m_H$(Gev) \hidewidth&\cr\tablerule
&&  .360  && .0000141 &&  2.707  &&  54.518 && 148.001 &\cr\tablerule
&&  .370  && .0000738 &&  2.656  &&  55.276 && 147.079 &\cr\tablerule
&&  .380  && .0001337 &&  2.602  &&  56.025 && 145.738 &\cr\tablerule
&&  .390  && .0001934 &&  2.543  &&  56.763 && 144.194 &\cr\tablerule
&&  .400  && .0002544 &&  2.479  &&  57.493 && 142.446 &\cr\tablerule
&&  .410  && .0003152 &&  2.408  &&  58.213 && 140.459 &\cr\tablerule
&&  .420  && .0003763 &&  2.328  &&  58.924 && 138.159 &\cr\tablerule
&&  .430  && .0004376 &&  2.235  &&  59.627 && 135.413 &\cr\tablerule
&&  .440  && .0004989 &&  2.120  &&  60.321 && 131.935 &\cr\tablerule
&&  .450  && .0005602 &&  1.958  &&  61.008 && 126.869 &\cr\tablerule
&&  .455  && .0005905 &&  1.823  &&  61.348 && 122.434 &\cr\tablerule
\noalign{\smallskip}
&\multispan9 \hfil\cr}}
\vskip1truecm

\vbox{\tabskip=0pt \offinterlineskip
\def\tablerule{\noalign{\hrule}}
\halign to400pt{\strut#& \vrule#\tabskip=1em plus2em&
   \hfil#& \vrule#& \hfil#& \vrule#& \hfil#&
   \vrule#& \hfill#& \vrule#& \hfil#& \vrule#
   \tabskip=0pt\cr\tablerule
&&\multispan9\hfil Table 2 \quad ($\overline G \ll \overline H $)
\hfil&\cr\tablerule
&&\omit\hidewidth $\overline T$\hidewidth&&
  \omit\hidewidth $\overline G$\hidewidth&&
  \omit\hidewidth $\overline H$\hidewidth&&
  \omit\hidewidth $m_t$(Gev) \hidewidth&&
  \omit\hidewidth $m_H$(Gev) \hidewidth&\cr\tablerule
&&  2.570  && .0005928 &&  1.807  &&  146.231 && 117.257 &\cr\tablerule
&&  2.575  && .0005332 &&  2.039  &&  146.374 && 125.012 &\cr\tablerule
&&  2.580  && .0004725 &&  2.173  &&  146.516 && 129.267 &\cr\tablerule
&&  2.585  && .0004110 &&  2.277  &&  146.658 && 132.461 &\cr\tablerule
&&  2.590  && .0003489 &&  2.365  &&  146.800 && 135.083 &\cr\tablerule
&&  2.595  && .0002862 &&  2.443  &&  146.942 && 137.332 &\cr\tablerule
&&  2.600  && .0002230 &&  2.513  &&  147.084 && 139.307 &\cr\tablerule
&&  2.605  && .0001592 &&  2.578  &&  147.225 && 141.062 &\cr\tablerule
&&  2.610  && .0000948 &&  2.638  &&  147.367 && 142.612 &\cr\tablerule
\noalign{\smallskip}
&\multispan9 \hfil\cr}}
\vskip1truecm

\vbox{\tabskip=0pt \offinterlineskip
\def\tablerule{\noalign{\hrule}}
\halign to400pt{\strut#& \vrule#\tabskip=1em plus2em&
   \hfil#& \vrule#& \hfil#& \vrule#& \hfil#&
   \vrule#& \hfill#& \vrule#& \hfil#& \vrule#
   \tabskip=0pt\cr\tablerule
&&\multispan9\hfil Table 3 \quad ($\overline G >\overline H$,\qquad
$\lambda <0 $) \hfil&\cr\tablerule
&&\omit\hidewidth $\overline T$\hidewidth&&
  \omit\hidewidth $\overline G$\hidewidth&&
  \omit\hidewidth $\overline H$\hidewidth&&
  \omit\hidewidth $m_t$(Gev) \hidewidth&&
  \omit\hidewidth $m_H$(Gev) \hidewidth&\cr\tablerule
&&  1.30  &&  184.169 &&  2.783  &&  104.069 && 1270.816 &\cr\tablerule
&&  1.50  &&  184.173 &&  2.789  &&  111.788 && 1208.710 &\cr\tablerule
&&  1.70  &&  184.174 &&  2.790  &&  119.007 && 1208.856 &\cr\tablerule
&&  1.90  &&  184.171 &&  2.786  &&  125.813 && 1208.153 &\cr\tablerule
&&  2.10  &&  184.164 &&  2.776  &&  132.269 && 1206.499 &\cr\tablerule
&&  2.30  &&  184.151 &&  2.759  &&  138.424 && 1203.787 &\cr\tablerule
&&  2.50  &&  184.134 &&  2.736  &&  144.317 && 1199.892 &\cr\tablerule
&&  2.61  &&  184.122 &&  2.719  &&  147.458 && 1197.194 &\cr\tablerule
\noalign{\smallskip}
&\multispan9 \hfil\cr}}
\end